\documentclass[prl,aps,showpacs,twocolumn,amsmath,amssymb]{revtex4-1}
\usepackage{epsfig}
\usepackage{amsmath}
\usepackage{amssymb}
\usepackage{multirow}
\usepackage{bibunits}
\usepackage{floatrow}
\defaultbibliography{nature}
\defaultbibliographystyle{unsrt}
\newcommand{\ke}{($\vec{k}, E$) }
\newcommand{\Siad}{Si$_{\mathrm{ad}}$}
\newcommand{\Agad}{Ag$_{\mathrm{ad}}$}
\newcommand{\Siado}{Si$^1_{\mathrm{ad}}$}
\newcommand{\Siadt}{Si$^2_{\mathrm{ad}}$}
\newcommand{\Sis}{Si$_{\mathrm{s}}$}
\newcommand{\Siso}{Si$^1_{\mathrm{s}}$}
\newcommand{\Sist}{Si$^2_{\mathrm{s}}$}
\newcommand{\ndot}{{\it nano-dot}}

\begin{document}
\begin{bibunit}[unsrt]

\title{Unveiling the {\it Penta}-Silicene nature of perfectly aligned single and double 
strand Si-nanoribbons on Ag(110)}

\author{Jorge I.~Cerd\'a$^{1}$}
\email{jcerda@icmm.csic.es}
\author{Jagoda S\l awi\'{n}ska$^1$}
\author{Guy Le~Lay$^{2}$}
\email{guy.lelay@univ-amu.fr}
\author{Antonela C.~Marele$^3$}
\author{Jos\'e M.~G\'omez-Rodr\'iguez$^{3,4}$}
\author{Mar\'ia .E.~D\'avila$^1$}
\affiliation{1. Instituto de Ciencia de Materiales de Madrid, ICMM-CSIC,
Cantoblanco, 28049 Madrid, Spain.}
\affiliation{2. Aix Marseille Universit\'e, CNRS, PIIM UMR 7345, 13397, Marseille, France}
\affiliation{3. Departamento de F\'{i}sica de la Materia Condensada, Universidad Aut\'onoma de Madrid, E-28049 Madrid, Spain.}
\affiliation{4. Condensed Matter Physics Center (IFIMAC), Universidad Aut\'onoma de Madrid, E-28049 Madrid, Spain.}
%\email[]{}

\date{\today}

%\begin{abstract}
%We reveal through extensive DFT calculations and STM simulations, confronted to
%key experimental facts, the hidden penta-silicene nature of single and double 
%strand chiral Si nanoribbons perfectly aligned on Ag(110) surfaces whose
%structure has remained elusive for over a decade since their discovery
%in 2005. After tracing the structures found at low coverages...
%{\it existence of penta-silicene, a recently conjectured novel pentagonal silicon 
%allotrope, which remained unveiled for 11 years, and which materializes a 
%paradigmatic shift from normal hexagonal silicene.}
%\end{abstract}

\maketitle

%\section{Introduction}
{\bf From the simplest cyclopentane ring and its numerous organic derivates to their
common appearance in extended geometries such as edges or defects in graphene, pentagons
are frequently encountered motifs in carbon related systems. Even a {\it penta}-graphene
Cairo-type two dimensional structure has been proposed as a purely pentagonal C
allotrope with outstanding properties competing with those of graphene~\cite{cpenta}.
Conversely, pentagonal Si motifs are hardly found in nature. Despite the
large effort devoted to design Si-based structures analogous to those of
carbon, the existence of Si pentagonal rings has only been reported in
complex clathrate bulk phases~\cite{calthrates}. Several theoretical studies
have hypothesized stable Si pentagonal structures either in the 
form of one-dimensional (1D) nanotubes~\cite{siwires1,siwires2} or 
at the reconstructed edges of {\it silicene} nanoribbons~\cite{silicene,silicene1} or 
even as hydrogenated {\it penta-silicene}~\cite{pentasilicene} or
highly corrugated fivefold coordinated {\it siliconeet}~\cite{grunberg} 2D sheets,
the latter recognized as a topological insulator~\cite{heine}.
However, to date none of them have yet been synthesized.
In the present work we unveil, via extended Density functional theory (DFT)~\cite{siesta} calculations and 
Scanning tunneling microscopy (STM) simulations~\cite{green,loit}, the atomic structure of 1D Si nano-ribbons grown on the 
Ag(110) surface. Our analysis reveals that this system constitutes the first 
experimental evidence of a silicon phase solely comprising pentagonal rings.}

\begin{figure}[h]
\includegraphics[width=75mm]{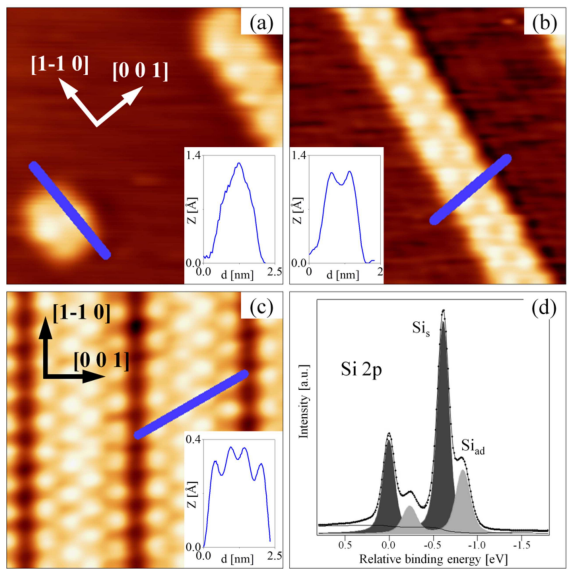}
\caption{(Color online) $5.3\times5.3$~nm$^{2}$ STM images of Si nanostructures
          on Ag(110). (a) a Si \ndot, (b) a Si SNR and (c) an extended Si DNR
          phase.
          The insets show profiles along the solid lines passing over the 
          maxima in the images. Tunneling
          parameters: (a) -1.5~V, 2.4~nA, (b) -1.8~V, 1.2~nA and (c) 1.3~V, 1.1~nA. 
          (d) Si-$2p$ core level photoemission spectra recorded at normal 
          emission and at 135.8~eV photon energy for the Si DNRs 
          structure.
         \label{exp}}
\end{figure}

%\section{Results}
Since their discovery in 2005~\cite{Leandri2005}
%and despite the numerous techniques
%employed towards their characterization
the atomic structure of Si
nano-ribbons (NRs) on Ag(110) has remained elusive and strongly 
disputed~\cite{chinese1,Leandri2005,ronci2010,Lian2012,Borensztein2014,Lagarde2016,arpes1,colonna,chinese}.
Figure~\ref{exp} presents a summary of Si NRs measured with
STM. 
The structures were obtained after Si sublimation onto a clean
Ag(110) surface at RT. Panels (a) and (b) correspond to a low Si 
coverage image with an isolated \ndot\ structure and a 
{\it single strand}
NR (SNR) 0.8~nm wide running along the $[1\bar{1}0]$ direction with a 
2$\times$ periodicity. The SNR topography consists of alternating
protrusions at each side of the strand with a glide plane.
At higher coverages and after a mild annealing, a dense and highly ordered 
phase is formed (panel (c)) consisting of {\it double strand} NRs (DNRs) with a 
5$\times$ periodicity along the [001] direction again exhibiting a
glide plane along the center of each DNR.
The images are in perfect accord with previous works~\cite{Leandri2005,ronci2010,colonna,chinese}. 
Further key information on the system is provided by the high-resolution
Si-$2p$ core level photoemission spectrum for the DNRs displayed in
Figure~\ref{exp}(d) --that for the SNRs is almost identical~\cite{Davila2012}.
The spectrum can be accurately fitted with only
two (spin-orbit splited) components having an intensity ratio of 
roughly 2:1. Furthermore, previous Angular Resolved Photoemission (ARPES) 
experiments~\cite{Paola2008}
assigned the larger and smaller components to subsurface
\Sis\ and surface \Siad\ atoms, respectively, indicating that the NRs 
comprise two different types of Si atoms, with twice as many \Sis\  as \Siad.

\begin{figure}[h]
\includegraphics[width=\columnwidth]{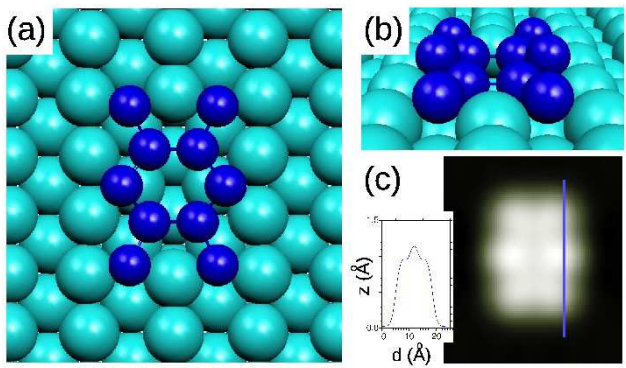}
\caption{
(a-b) Top and perspective views of the \ndot\ structure.
(c) Simulated STM topographic image and line profile
along the solid line.
     \label{nano}}
\end{figure}

We first focus on the \ndot\ 
shown in Fig.~\ref{exp}(a), as it may be regarded as the precursor structure
for the formation of the extended NRs. The \ndot\ exhibits a local
{\it pmm} symmetry with two bright protrusions aligned along the [001] 
direction, each of them having two adjacent dimmer features along the
$[1\bar{1}0]$ direction. After considering a large variety of trial models
(see 'Extended Data' Fig~\ref{nano_all}) we found that only one, shown in 
Figure~\ref{nano}, correctly reproduces the experimental image both
in terms of aspect and overall corrugation.
It consists of a ten atom Si cluster located in a double silver vacancy
generated by removing two adjacent top row silver atoms.
There are four symmetry equivalent \Sis\ atoms residing deeper in the 
vacancy, two \Siado\ in the middle which lean towards short silver 
bridge sites and four outer \Siadt\ residing at
long bridge sites. The formers lie 0.8~\AA\ above the top Ag atoms and 
are not resolved in the STM image, while the \Siado\ and \Siadt\
protrude out of the surface by 1.4 and 1.1~\AA\ thus leading to the six 
bump structure in the simulated image with the \Siado\ at the center 
appearing brighter. 
%{\it In order to confirm the creation of the double 
%vacancy, an analogous structure was additionally considered after 
%replacing the two \Siado\ by silver adatoms. Such model would be consistent 
%with a double Ag-Si site exchange mechanism without requiring the diffusion of 
%the metal atoms across long distances over the surface. However, the associated
%STM image, shown in Fig.~\ref{nano}(d), is in clear disagreement with the 
%experimental one.}
Therefore, although the \ndot\ shows marked differences with
respect to the extended NRs, its structure already accounts for the presence of two 
distinct types of Si atoms at the surface (\Sis\ and \Siad). Furthermore,
it reveals the tendency of the Ag(110) surface 
upon Si adsorption to remove top row silver atoms 
(i.e. the initial stage in the creation of a missing row (MR))
and incorporate Si nanostructures in the troughs.
%instead of a creating a 1:1 substitutional alloyed surface 
%often found for foreign atoms on (110) metal surfaces~\cite{agsb,nimn}. 

\begin{figure}[h]
\includegraphics[scale=0.55]{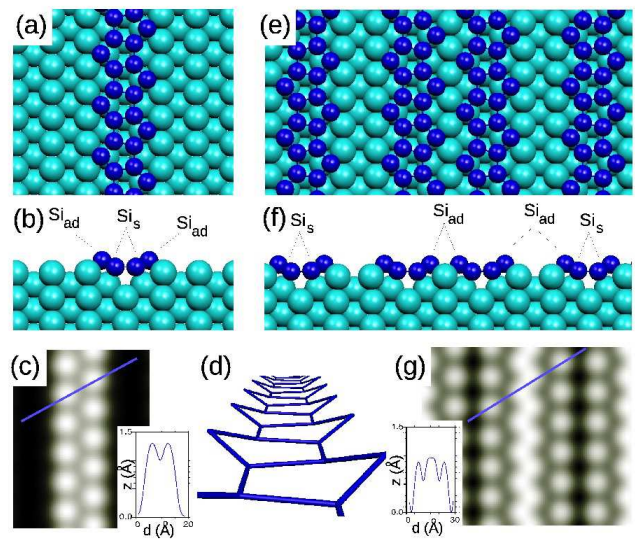}
\caption{Optimized geometry of the pentagonal missing row (P-MR) model
         (a-c) Top, side and simulated topographic STM image for the SNR phase.
         (d) Perspective view of a penta-silicene strand without the silver
         surface.
         (e-g) Top, side and simulated topographic STM image for the DNR array.
         Insets in (c) and (g) show line profiles along the blue lines 
         indicated in the topographic maps.
         All STM simulations employed a sharp Si ended tip 
         apex and set points $V=-0.2$~V and $I=1$~nA.
         \label{pmr}}
\end{figure}

\begin{figure*}[t]
 \includegraphics[width=\textwidth]{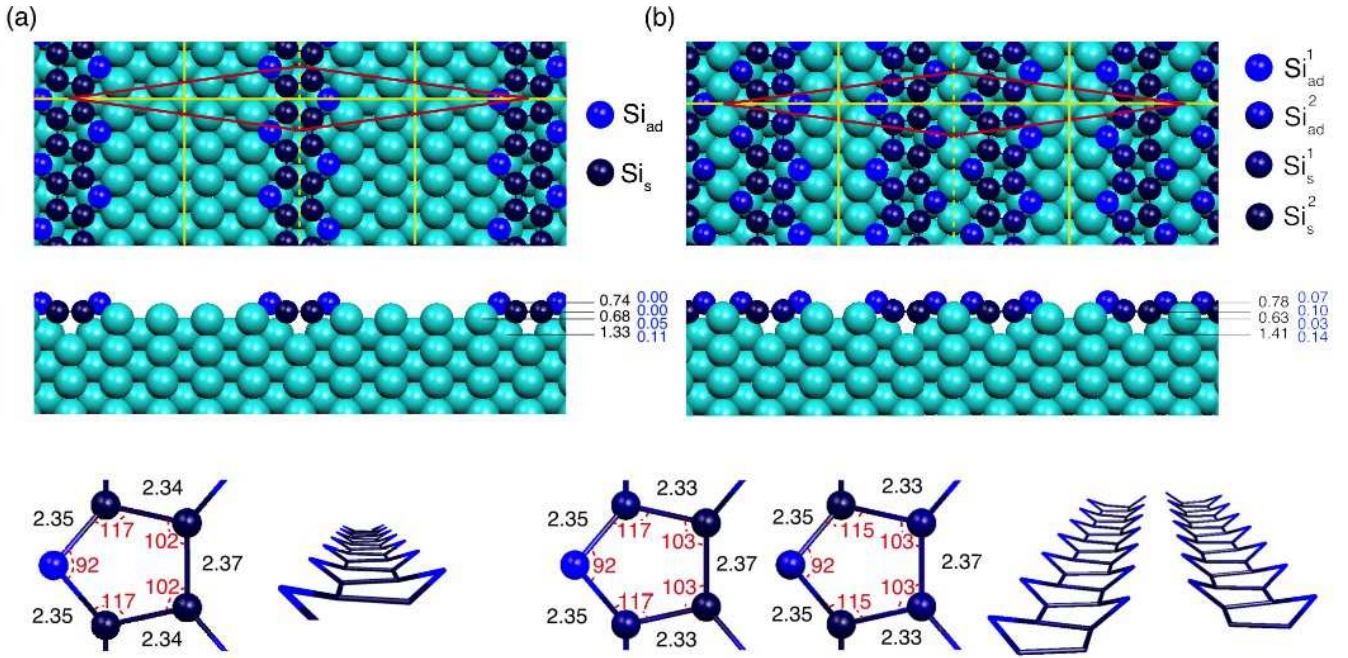}
 \caption{{\bf Geometry of the P-MR model:}
          P-MR/Ag(110) optimized structures for the Si SNRs (a) and the DNRs (b).
The Si atoms have been colored according to the symmetry inequivalent group they
belong (see legends).
Top panels: top views with the $c(10\times2)$ supercell indicated by the dark
rhombus while green thick and dashed lines correspond to mirror and glide symmetry planes,
respectively. The overall symmetry for both models is $cmm$.
Middle panels: side views with the normal averaged distances between the \Siad,
\Sis\ and the two first silver layers indicated by black numbers, and the buckling
within each group of atoms given by blue numbers (all distances in \AA).
Bottom panels: Zoom in of the pentagonal rings including the Si-Si bond
distances (in \AA) and bond angles (in red), and perspective views of the 1D
pentagonal structures. 
In Table~\ref{nn} we additionally provide the
relative $z$-coordinate of the Si atoms and their nearest neighbor distances
to the metal atoms. In the DNR phase
the loss of the local glide plane within each NR makes the two \Siad\
at each side of the pentagonal chains inequivalent, with the outer ones
(\Siado) lying 0.06~\AA\ above the inner ones (\Siadt) while
their lateral distance to the top row bridge site is 0.1~\AA\ smaller
for the formers. Both trends
may be explained from the symmetry constrain on the top row silver atoms
along the glide plane, as they cannot shift laterally, as well as by
certain Ag-mediated repulsion between the \Siadt\ in adjacent NRs
(now each top silver atom makes two bonds with the \Siadt)
that shifts the pentagonal rings away from each other by 0.2~\AA.
Concerning the substrate MR reconstruction
there is a 0.1~\AA\ lateral shift of the top row atoms away from
the troughs in order to better accommodate the Si NRs.
Finally we recall that the fact that
the four \Siad\ in the DNR are not colinear is a key factor in determining the
aspect (arrangement of the protrusions) in the STM images.
\label{geom}}
\end{figure*}

\begin{figure*}
\includegraphics[scale=0.6]{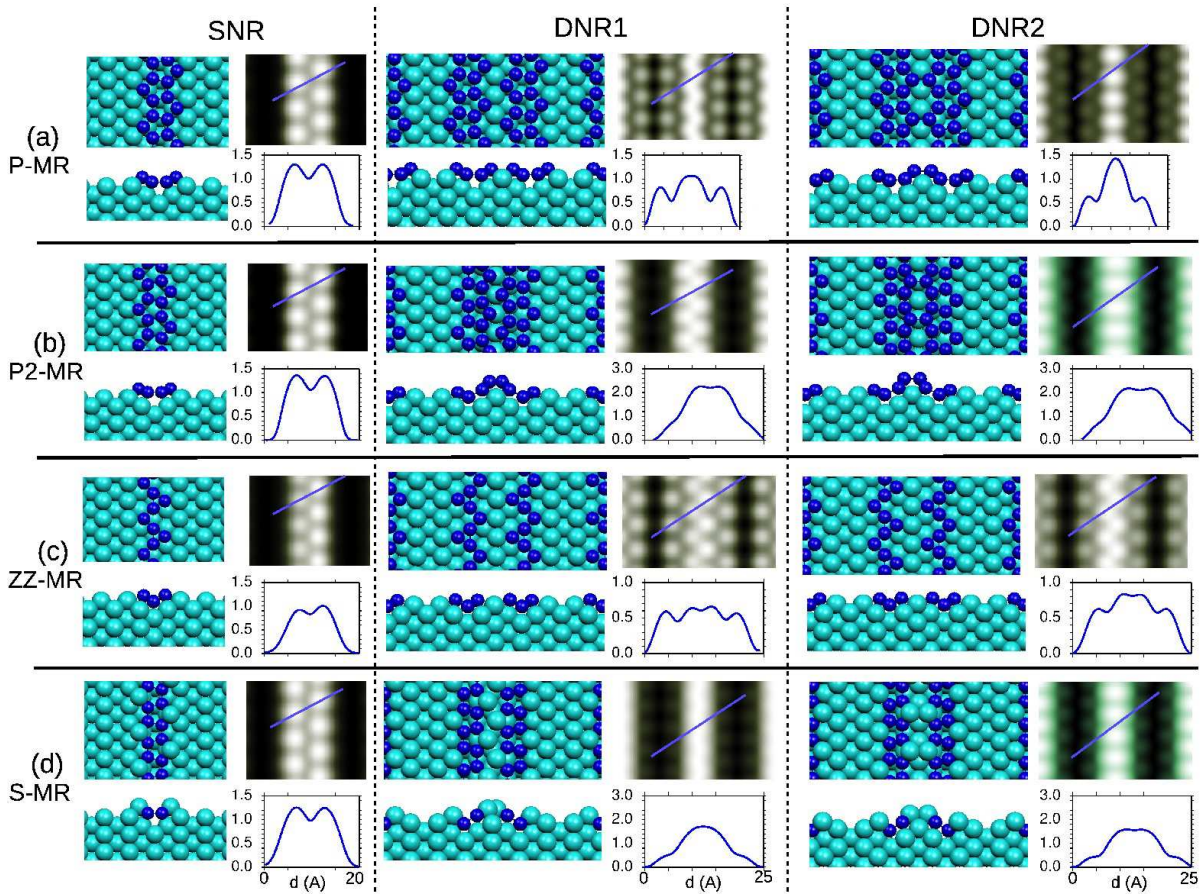}
\caption{{\bf Trial MR-NR models:}
          Optimized geometries, STM topographic maps and line profiles for
the most relevant MR-based NR structures considered in this work
(we omit the tens of models tested based on an unreconstructed Ag(110) surface since
they systematically relaxed towards geometries incompatible with
the experimental STM images).
Left panels correspond to the low coverage SNRs, and center and right
columns to the high coverage DNR arrangements following a $-LL-RR-LL$ (DNR1)
and a $-LR-RL-LR-$ (DNR2) sequence among the enantiomers, respectively.
(a) P-MR model already shown in Figure~\ref{pmr} together with the alternative
DNR2 arrangement whereby the \Siad\ become aligned between the adjacent 
rings and dimerize on top of the short bridge sites leading to
a broad ribbon made up of elongated octagons at the center decorated by
pentagons at the edges.
(b) P2-MR model similar to the P-MR one but with the \Siad\
leaning towards top sites.
(c) zig-zag missing row model (ZZ-MR) proposed in Ref.~\cite{colonna}.
The motif in this pattern consists of only two \Sis\ atoms residing in the MR
troughs, each of them bonded to another two \Siad s
which lean towards the Ag atoms at the top row and protrude out of the surface.
(d) a substitutional model (S-MR) where the extracted silver
atoms attach to the \Sis\ located in the MR troughs (i.e. equivalent to the
P-MR but with the \Siad\ replaced by \Agad).
Such model would be consistent with a site exchange mechanism between two Si 
and one silver top row atom without the need for diffusion of the latter across
long distances over the surface. 
All SNRs yield STM images highly reminiscent of the experimental one 
(Fig.~\ref{exp}(b)), with the only exception of the ZZ-MR structure, which 
shows an asymmetry in the protrusions due to the lack of a glide plane.
However, in both DNR arrangements for the P2-MR and S-MR models as well as
in the P-MR DNR2 structure, the adatoms (\Siad\ or \Agad) are raised above 
the silver top row by large distances in order to establish bonds among the 
adjacent NRs and the corresponding STM images deviate substantially from the 
high coverage experimental one (Fig.~\ref{exp}(c)) as their aspect is now 
dominated by these high lying Si/Ag atoms. Therefore, these structures may be
discarded based on their STM topography.
The ZZ-MR DNRs, on the other hand, present a nice correspondence with
Fig.~\ref{exp}(c), specially the DNR2 arrangement which exhibits a glide
plane symmetry. Anyhow, this model may be ruled out as well based on energetic
arguments (see Table~\ref{eads} and Fig.~\ref{phasediag}).}
\label{models}
\end{figure*}

%{\it Pentagonal nature of the NRs--} 
Inspired by the \ndot\ Ag di-vacancy structure and by recent
grazing incidence X-ray diffraction (XRD) measurements~\cite{mr} pointing towards the
existence of a MR reconstruction along the $[1\bar{1}0]$ direction
of the Ag surface, we considered several trial structures for
the SNRs by placing Si atoms in the MR troughs (\Sis) and next
adding further adatoms (\Siad) on top, while maintaining a 2:1 
concentration ratio between the two.
Figures~\ref{pmr}(a-b) show top and side views of the optimized geometry
for the SNRs after testing several trial models
(see 'Extended Data' Fig.~\ref{models}).
It involves a MR and six Si atoms per cell.
The new paradigm is
the arrangement of the Si atoms into pentagonal rings running along the MR 
and alternating their orientation (we denote it as the P-MR model).
Despite no symmetry restriction was imposed, the relaxed P-MR SNR belongs
to the $cmm$ group presenting two mirror planes plus
an additional glide plane along the MR troughs (see 'Extended Data' 
Fig.~\ref{geom} for a detailed description).
Apart from a considerable buckling of 0.7~\AA\ between the lower Si atoms
residing in the MR troughs (\Sis) and the higher ones (\Siad) leaning
towards short bridge sites at the top silver row,
the pentagonal ring may be considered as rather perfect, with a very
small dispersion in the Si-Si distances ($2.35-2.37$~\AA) and bond angles
ranging between $92^\circ-117^\circ$; that is, all close to the $108^\circ$
in a regular pentagon.
The associated STM image and line profile, panel (d), show (symmetry) 
equivalent protrusions 1.3~\AA\ high at each side of
the strand, in perfect agreement with the experimental image. 
Still, since different models may yield
similar STM images, a more conclusive gauge to discriminate among them is
to examine their relative formation energies. In this respect,
the energetic stability of the P-MR structure is 
far better ($\sim 0.1$~eV/Si) than all other SNR models considered 
(see section 'Methods' and 'Extended Data' Fig.~\ref{phasediag}).

Within the pentagonal model the DNR structure may be naturally generated
by placing two SNRs within a $c(10\times2)$ cell. However, since the P-MR SNRs are chiral,
adjacent pentagonal rings may be placed with the same or with different 
handedness, leading to two possible arrangements among
the enantiomers.
Figures~\ref{pmr}(e-g) display the optimized geometry and
simulated STM topography for the most stable (by 0.03~eV/Si) P-MR DNR
configuration.
The pentagonal structure in each NR is essentially preserved, the main 
difference with respect to the SNRs being the loss of the glide plane
along the MR troughs replaced by a new one along the top 
silver row between adjacent SNRs. There is a slight
repulsion between the NRs which shifts them away from each other
by around 0.2~\AA. As a result, the \Siad\ at the outer edges of the DNR
end up lying 0.07~\AA\ higher than the inner ones making the alternating 
pentagons along each strand not strictly equivalent anymore.
In the simulated STM image the outer maxima appear dimmer
than the inner ones by 0.1~\AA, which adopt a zig-zag aspect.
The inversion in their relative corrugations is due to the proximity  
between the inner Si adatoms ($\sim4$~\AA) compared to the almost 6~\AA\
distance between the inner and outer ones, so that the bumps of the
formers overlap and lead to brighther maxima.
%and merge into a single broader and higher maximum.
All these features are in accordance with the experimental profiles shown 
in Fig.~\ref{exp}(c).
%The alternative $-RL-LR-RL-$ arrangement leads to an STM
%image incompatible with the experimental one (see {\it SM~\#4}) and is less
%stable (by 0.03~eV per Si atom) than the P-MR DNR phase.
%{\it The alternative $-RL-LR-RL-$ arrangement, whereby the \Siad\ become aligned 
%between the adjacent rings and dimmerize on top of the short bridge sites relaxes
%towards a broad ribbon made up of elongated octagons at the center decorated by 
%pentagons at the edges. 
%Since the \Sis\ remain 
%tightly binded to the subsurface silver atoms, there is and increase in
%the overall pentagon's corrugation of up to 1.3~\AA. 
%The corresponding STM image shows horizontally elongated 
%maxima on top of the dimers in clear discrepancy with the experimental images
%for the DNRs.
%Nevertheless, such an appealing structure leads to an STM
%image incompatible with the experimental one (see {\it SM~\#4}) and is less
%stable (by 0.03~eV per Si atom) than the P-MR DNR phase (see {\it SM~\#5}) 
%indicating that the energy penalty paid upon increasing the
%corrugation in the latter ($sp^2$ to $sp^3$ transition) is not compensated by the 
%gain associated to the dimmerization.}
In fact, the P-MR DNR structure is the most stable among all other NR models
considered for a wide range of Si chemical potentials ranging from Si-poor to
-rich conditions (see 'Extended Data' Fig.~\ref{phasediag}).

\begin{figure}[h]
\includegraphics[width=\columnwidth]{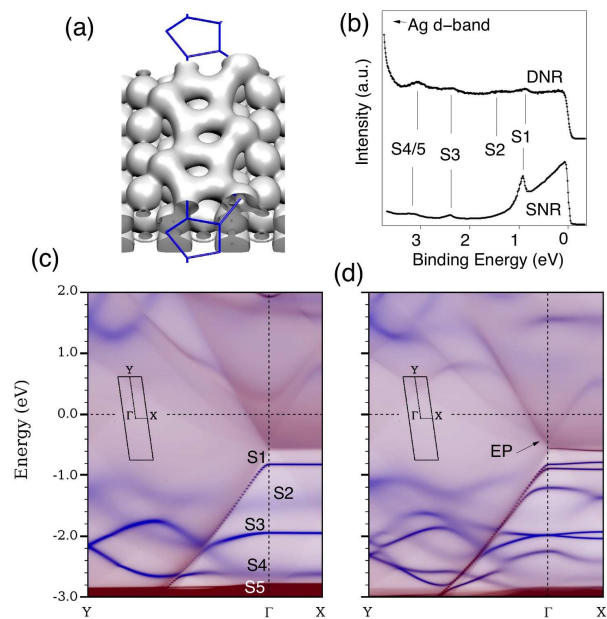}
\caption{Electronic structure of the P-MR model. 
(a) Charge density isosurface for the SNRs with blue sticks indicating 
the Si pentagons. (b) Energy distribution curves around the $X$ point 
for the SNRs acquired at 78~eV photon energy 
(adapted from Ref~\cite{arpes2}) and for the DNRs at 75~eV.
(c-d) PDOS\ke projected on the Si (blue) and Ag (red) atoms along
the $Y-\Gamma-X$ $k$-path (see insets) for the SNRs and DNRs, respectively.
     \label{electronic}}
\end{figure}

%{\it The Si coverage of $\Theta_{Si}=1.2$~ML in the P-MR DNRs is somewhat larger
%than that deduced before~\cite{mr,colonna}, but we recall that a precise 
%determination of the exact coverage is actually extremely difficult and 
%generally highly questionable.}
%{\it Band structure--}
Figure~\ref{electronic} presents a summary of the electronic properties of the 
P-MR structure. Panel (a) shows an isosurface of the total electronic 
density for the SNRs. The \Sis\ atoms
in the pentagonal rings are clearly linked through an $sp^2$ type bonding 
(three bonds each) while the \Siad, due to the buckling, show a 
distorted $sp^3$ type tetrahedral arrangement making bonds with two \Sis\ 
as well as with the adjacent short bridge silver atoms in the top row.
Panel (b) displays ARPES spectra for the SNR and DNR phases. 
Both energy distribution curves reveal Si-related peaks previously 
attributed to {\it quantum well states} (QWS) originating from the 
lateral confinement within the NRs. 

%Their intensity, however, was found strongly dependent on the 
%orientation of the incident light, the recorded photoelectron emission angle 
%as well as the photon energy and the polarization vector of the light.
For the SNRs three states are observed at -1.0, -2.4 and 
-3.1~eV binding energy (BE), while for the DNRs
one further peak is identified at -1.4~eV. 
The computed (semi-infinite) surface band structures projected on the Si 
pentagons (blue) and the silver MR surface (red) are superimposed in
panels (c) and (d) for the SNRs and DNRs, respectively. 
Overall, within the expected DFT accuracy and 
experimental resolution, the maps satisfactorily reproduce
the experimental spectra. 
At $\Gamma$ the SNRs present two sharp intense Si bands below the
Fermi level (S1 and S3) and faint (broader)
features arising from two almost degenerate bands (S4 and S5) and a
dimmer state (S2).
As expected, they are almost flat along $\Gamma-X$ while along
$\Gamma-Y$ they present an appreciable dispersion and finally merge into two
degenerate states at the high symmetry $Y$ point. The orbital character of
the S2-S5 bands is mainly $p_{xy}$ and may thus be assigned to localized 
$sp^2$ planar bonds. Conversely, band S1 is fully dominated by the
Si$_s-p_z$ states ($\pi$-band) and shows a strong downward dispersion along
$\Gamma-Y$ due to hybridization with the metal $sp$ bands. Similarly,
faint dispersive bands of mainly $p_z$ character hybridizing with the metal
appear in the empty states region.
The electronic structure for the DNRs is similar to that of the SNRs, except 
that the number of Si bands is doubled and most of them become splited and
shifted due to the interaction between adjacent SNRs. Noteworthy is
the appearance of an electron pocket (EP) at $\Gamma$ associated to a 
parabolic Si-$p_z$ band with onset at -0.5~eV.

\begin{figure}[b]
\includegraphics[scale=0.5]{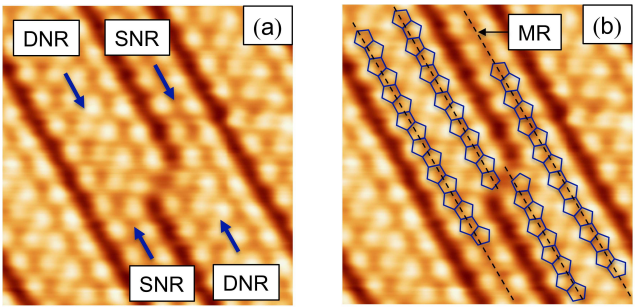}
\caption{{\bf Dislocation defects between NRs.}  (a) Experimental STM image 
     showing a dislocation within the array of DNRs. The image size is 
     $6.2\times6.2$~nm$^{2}$. The sample bias voltage is 0.9~V and the 
     tunnel current 0.8~nA. (b) Same as (a) after superimposing the
     Si pentagons (blue) and the MRs (dashed solid lines).
     The upper and lower truncated MRs at the 
     center are shifted from each other by one Ag lattice parameter, leading
     to a DNR-SNR arrangement at the top of the image and a SNR-DNR at the
     bottom.}
\label{defect}
\end{figure}

To conclude, we have solved the long debated structure of silicon nano-ribbons
on Ag(110), finding an unprecedented 1D {\it penta-silicene} phase which
consists of adjacent inverted pentagons stabilized within the MR troughs.
The model is in accordance with most of previous experimental results for this
system: it involves a MR reconstruction as deduced from XRD~\cite{mr}, 
comprises two types of Si atoms with a ratio 2:1 between the \Sis\ and \Siad\
concentrations as seen by photoemission, accurately matches the STM topographs 
also explaining dislocation defects between NRs (see 'Extended Data' 
Fig.~\ref{defect}) and accounts for the QWS measured by ARPES.
We have also determined the quasi-hexagonal geometry of a
Si \ndot\ inside a silver-divacancy. This precusor structure for the
NRs can be considered as the limiting process for expelling surface 
Ag atoms in order to create a missing row along which the Si pentagons 
can develop.
We are convinced that the discovery of this novel silicon allotrope
will promote the synthesis of analogous exotic Si phases on alternative 
templates with promising properties~\cite{grunberg}.

%\section{Bibliography}
%\bibliography{nature}

\putbib[nature]
\end{bibunit}

%\section{Acknowledgements}
\vspace{2 cm}
{\bf Acknowledgements:}
This work has been funded by the Spanish MINECO under contract Nos.
MAT2013-47878-C2-R, MAT2015-66888-C3-1R, CSD2010-00024, MAT2013-41636-P, 
AYA2012-39832-C02-01/02 and ESP2015-67842-P.

%\section{Author contributions}
{\bf Author contributions:}
J.I.C. and J.S. performed all the theoretical calculations.
A.C.M., M.E.D., and J.M.G.R. performed all the STM experiments. M.E.D. and G.L.L. performed the ARPES measurements.
J.I.C. and M.E.D. conceived most of the
novel model structures tested. J.I.C. and G.L.L. wrote the manuscript. 
All authors contributed to the manuscript and figure preparation.

\begin{bibunit}

\section{Methods}

%\subsection{Experimental}
{\bf Experimental}

For both types of prepared structures (isolated Si SNRs or ordered DNRs),
the same procedure has been used for sample preparation: i.e. the 
Ag(110) substrate was cleaned in the Ultra-high vacuum (UHV) chambers (base pressure:
9$\times10^{-11}$~mbar) by repeated sputtering of Ar$^+$ ions and subsequent 
annealing of the substrate at 750~K, while keeping the pressure below 
3$\times10^{-10}$~mbar during heating. Si was evaporated at a rate
of 0.03~ML/min from a silicon source in order to form the NRs. The Ag 
substrate was kept at room temperature RT to form the isolated
SNR 0.8~nm wide, while a mild heating of
the Ag substrate at 443~K allows the formation of
an ordered grating DNR 1.6~nm wide~\cite{Davila2012}.

STM measurements were done with a home-made variable temperature UHV 
STM~\cite{Custance2003}. All STM data were measured and processed with the 
WSxM software \cite{Horcas2007}. High-Resolution Photoelectron Spectroscopy 
(HRPEs) experiments of the shallow Si-$2p$ core-levels and of the valence states,
were carried out to probe, comparatively, the structure and the electronic 
properties of those nanostructures. The ARPES experiments were carried out at the 
I511 beamline of the Swedish Synchrotron Facility MAX-LAB in Sweden.
The end station is equipped with a Scienta R4000 electron 
spectrometer rotatable around the propagation direction of the synchrotron 
light. It also houses low energy electron diffraction (LEED) and sputter
cleaning set-ups. Further details on the beam line are given in Ref.~\cite{Denecke1999}.
In all the photoemission spectra the binding energy is referenced
to the Fermi level. The total experimental resolution for core level and
valence band (VB) spectra were 30~meV (h$\nu$=135.8~eV for Si-$2p$) and 20~meV
(h$\nu$=75~eV for the VB), respectively. 
A least-square fitting procedure was used to analyze the core-levels, with two doublets, 
each with a spin-orbit splitting of $610\pm5$~meV and a branching ratio of 0.42. 
The Si-$2p$ core level collected at  normal emission  is dominated by the \Sis\  component. 
Its full width at half maximum (FWHM) is only 68~meV while the energy 
difference between the two \Sis\ and \Siad\ components is 0.22~eV.

{\bf Theory}

All calculations have been carried out at the 
{\it ab initio} level within the Density Functional Theory (DFT) 
employing the SIESTA-GREEN package~\cite{siesta,green}. 
For the exchange-correlation (XC)
interaction we considered both the Local Density~\cite{ca} (LDA)
as well as the Generalized Gradient~\cite{pbe} (GGA) approximations.
The atomic orbital (AO) basis set consisted of Double-Zeta
Polarized (DZP) numerical orbitals strictly localized after setting a
confinement energy of 100~meV in the basis set generation process.
Real space three-center integrals were computed over 3D-grids with
a resolution equivalent to a 700~Rydbergs mesh cut-off. Brillouin zone (BZ)
integration was performed over $k$-supercells of around (20$\times$28) relative
to the Ag-(1$\times$1) lattice while the temperature $kT$ in the Fermi-Dirac
distribution was set to 100~meV.

All considered Si-NR-Ag(110) structures were
relaxed employing two-dimensional periodic slabs involving nine metal layers
with the NR adsorbed at the upper side of the slab. A $c(10\times2)$
supercell was employed for both the SNR and DNR structures.
In all cases, the Si
atoms and the first three metallic layers were allowed to relax until forces
were below 0.02~eV/\AA\ while the rest of silver layers were held fixed to
their bulk positions (for which we used our LDA (GGA) optimized lattice
constant of 4.07~\AA\ (4.15~\AA), slightly smaller (larger) than the
4.09~\AA\ experimental value).

\begin{figure*}
\includegraphics[scale=0.45]{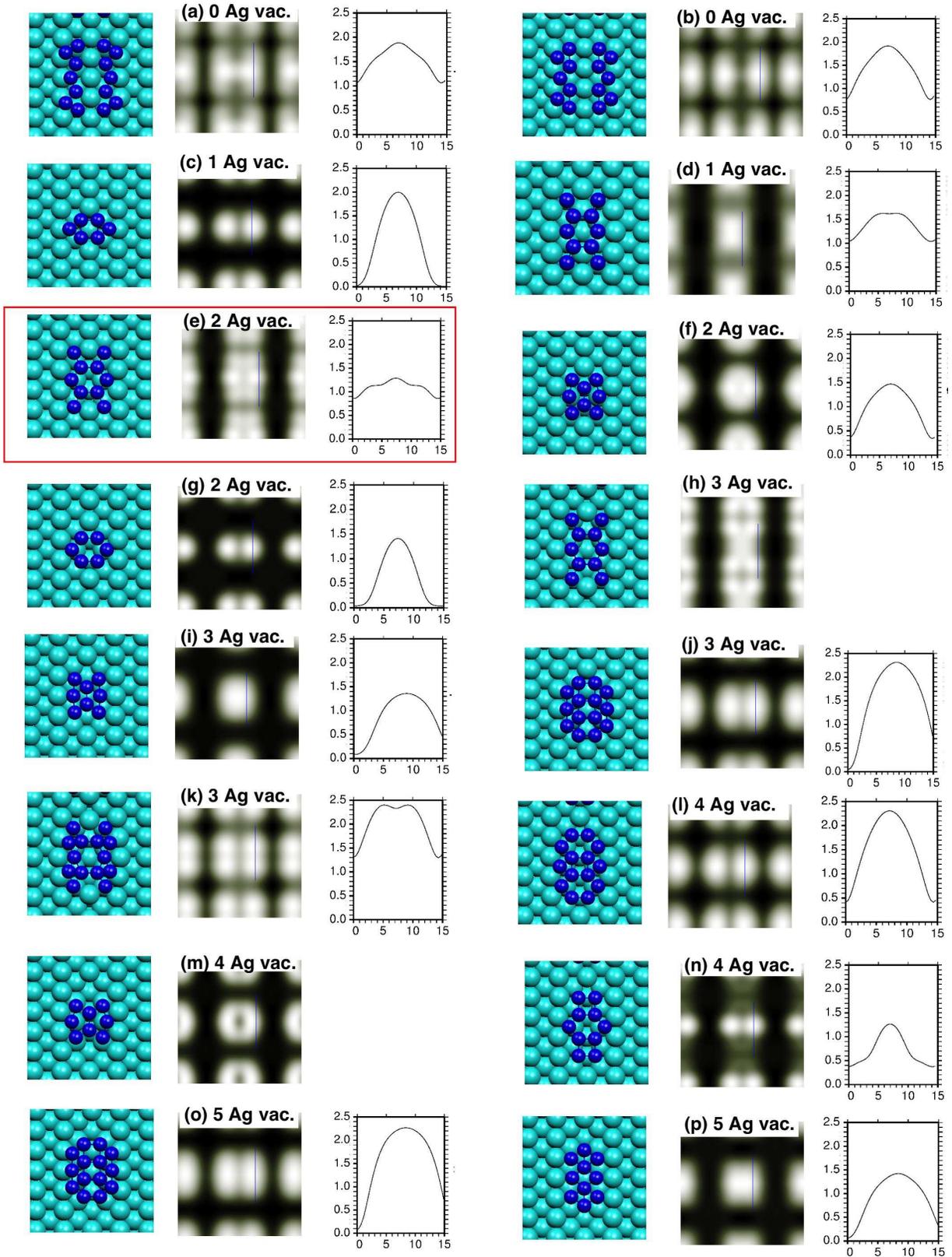}
\caption{{\bf Trial nano-dot models:}
Summary of the most relevant \ndot\ trial structures studied in this work employing
a reduced (4$\times$5) or (4$\times$6) supercell.
Optimized geometries (top view), STM simulations and line profiles
along the vertical blue lines shown in the maps for all \ndot\ models tested.
The models are organized from (a-p)
with increasing number of silver vacancies, ranging from zero up to five.
The structure that best matches
the experimental image and line profile given in Fig.~\ref{nano}(a) is clearly
case (d), involving two Ag vacancies and ten Si atoms. The simulated image
correctly captures the two large bumps at the center and the dimmer maxima
(elbows)
above and below them. Furthermore, the associated profile is the only one that
resolves the three maxima with just a few tenths of Angstrom difference between
the center and the satellite ones, in perfect agreement with the experiment.
\label{nano_all}}
\end{figure*}

For the \ndot\ calculations, and given that a larger
unit cell is required to simulate its isolated geometry, the atomic relaxations
of all the trial models (see Fig.~\ref{nano_all}) were carried out for (4$\times$5)
or (4$\times$6) supercells. STM topographic
images were next computed for all relaxed structures after
recomputing the slab Hamiltonians with highly-extended AOs for the surface
atoms. Once the correct structure was identified (see Fig.~\ref{nano_all}), 
we further optimized it increasing the unit cell to a $(6\times10)$ to remove
any overlaps between image cells (see Fig.~\ref{nano} in the main text).

{\it Band structure--} 
In order to examine the surface band dispersion we computed $k$-resolved
surface projected density of states PDOS($k,E$) maps in a semi-infinite
geometry. To this end we stacked the Si-NR and first metallic layers on top of
an Ag(110) bulk-like semi-infinite block via Green's functions matching
techniques following the prescription detailed elsewhere~\cite{ysi2,loit}.
For this step we recomputed the slab's Hamiltonian employing highly
extended orbitals (confinement energy of just 10-20~meV)
for the Si and Ag surface atoms in the top two layers (this way the spatial
extension of the electronic density in the vacuum region is largely extended
and the calculation becomes more accurate.

\begin{figure}
\includegraphics[scale=0.4]{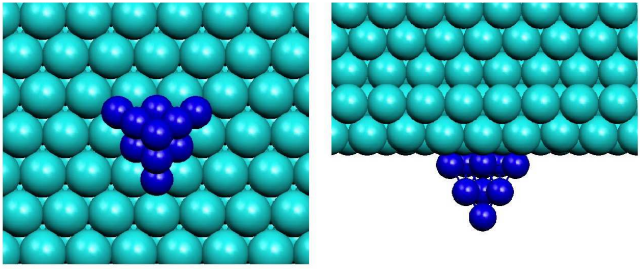}
\caption{{\bf Si$_{10}$-Ag(111) tip:}
         Bottom and side views of the Ag(111) tip terminated in a 10 Si atom
         pyramid employed for all STM simulations.}
\label{tip}
\end{figure}

{\it STM simulations--}
For the STM simulations we modeled the tip as an Ag(111) semi-infinite
block with a one-atom terminated pyramid made of ten Si atoms stacked below
acting as the apex (see Figure~\ref{tip}). Test calculations employing other
tips (e.g. clean Ag or clean W) did not yield any significant changes.
Highly extended orbitals were also employed
to describe the apex atoms thus reproducing better the expected exponential
decay of the current with the tip-sample normal distance $z_{tip}$. Tip-sample
AO interactions were computed at the DFT level employing a slab including the
Si NR on top of three silver layers as well as the Si tip apex.
The interactions (Hamiltonian matrix elements) were stored for
different relative tip-surface positions and next fitted to obtain
Slater-Koster parameters that allow a fast and accurate evaluation of
these interactions for any tip-sample relative position~\cite{loit}.
Our Green's
function based formalism to simulate STM images includes only the elastic
contribution to the current and assumes just one single tunneling process
across the STM interface; it has been extensively
described in previous works~\cite{green,loit}. Here we employed
an imaginary part of the energy of 20~meV which also corresponds to
the resolution used in the energy grid when integrating the transmission
coefficient over the bias window. We further assumed the so called wide band
limit (WBL) at the tip~\cite{loit} in order to alleviate the computational cost
and remove undesired tip electronic features. The images were computed at
different biases between -2 to +2~V scanning the entire unit cell
with a  lateral resolution of 0.4~\AA\ always assuming a fixed current of 1~nA.
Nevertheless, the aspect of the images hardly changed with the bias,
in accordance with most experimental results.

{\it Energetics--}
To establish the energetic hierarchy among different Si NR structures
we first computed their adsorption energies (per Si atom),
$E_{\mathrm{ads}}$, via the simple expression:
\begin{equation} \label{eq:ads}
E_{\mathrm{ads}} = \left(E_{\mathrm{tot}}(N_{Ag},N_{Si})-E_{\mathrm{surf}}(N_{Ag})-
    N_{\mathrm{Si}}E^0_{\mathrm{Si}}\right)/N_{\mathrm{Si}}
\end{equation}
where $N_{\mathrm{Si/Ag}}$ are the number of Si and Ag atoms in the
slab containing the NR and the Ag(110) surface,
$E_{\mathrm{tot}}(N_{Ag},N_{Si})$ refers to its total energy,
$E_{\mathrm{surf}}(N_{Ag})$ the energy of the clean Ag surface without the NRs
(but including any MRs) and
$E^0_{\mathrm{Si}}$ the energy of an isolated Si atom. In the low temperature
limit eq.~(\ref{eq:ads}) allows to discriminate between structures with
the same number of silver and Si atoms. 

However, a more correct approach to compare the NR's stabilities
between structures with different Si and Ag concentrations is to compute
their formation energies, $\gamma$, as a
function of the Si and Ag chemical potentials, $\mu_{Si/Ag}$. To this end, we
employ the standard low temperature expression for the grand-canonical
thermodynamic potential~\cite{phasediag}:
\begin{equation}
\Omega(\mu_{\mathrm{Si}}, \mu_{\mathrm{Ag}})=E_{\mathrm{tot}}(N_{\mathrm{Si}}, N_{\mathrm{Ag}})-N_{\mathrm{Si}}\mu_{\mathrm{Si}} - N_{\mathrm{Ag}}\mu_{\mathrm{Ag}}
\end{equation}

The chemical potentials may be obtained via $\mu_{Ag/Si}=E^{ref}_{Ag/Si}-
E^0_{Ag/Si}$, where $E^0$ corresponds to the total energy of the isolatad
atom and $E^{ref}$ to that of a reference
structure acting as a reservoir of Ag or Si atoms. Here we use the bulk $fcc$
phase for silver ($\mu^{LDA}_{Ag}=-4.67$~eV and $\mu^{GGA}_{Ag}=-3.60$~eV), while that
of Si is considered as a parameter (see below).
The NR's formation energy,
normalized to the Ag(110)-$(1\times1)$ surface unit cell area,
then takes the form:
\begin{equation}\label{eq:form}
\gamma=\frac{1}{N}[E_{tot}(N_{Si},N_{Ag})  - 
       N_{Ag} E^{ref}_{Ag} - N_{Si} \mu_{Si}] - \gamma^{sb}_{Ag}
\end{equation}
with $N=10$ because the same $c(10\times2$) was used for all NR structures and
$\gamma^{sb}$ accounts for the formation energy of
the unrelaxed surface at the bottom of the slab, which was obtained according to:
$\gamma^{sb}_{Ag}=\frac{1}{2}[E_{1\times1}(N_{Ag})- N_{Ag} E^{ref}_{Ag}]$
with $E_{1\times1}(N_{Ag})$ giving the total energy of an unrelaxed nine
layers thick Ag(110)-(1$\times$1) slab.

\begin{figure*}[h]
\includegraphics[width=\textwidth]{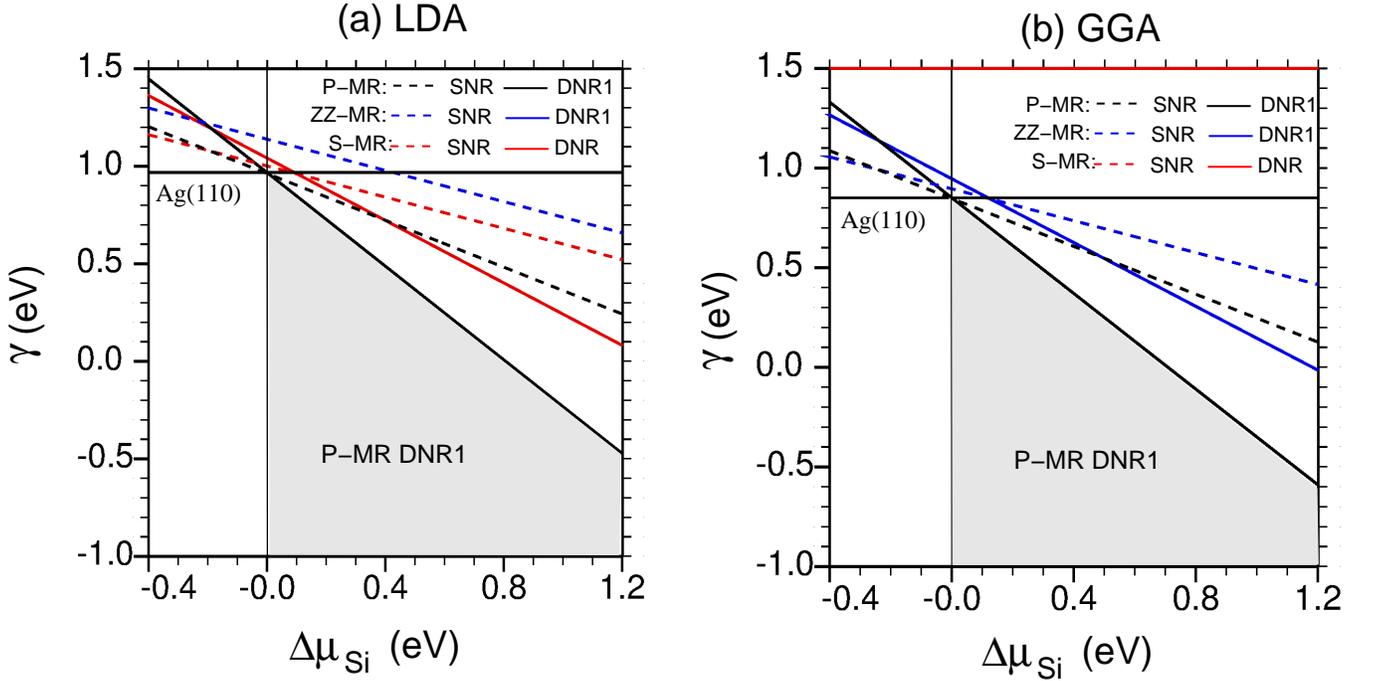}
\caption{{\bf Stability of the NR trial models:}
         Phase diagram for the different Si-NR-Ag(110) models studied in this
work under the (a) LDA and (b) GGA. Formation energies, normalized to the
Ag(110)-($1\times1$) unit cell,
are plotted as a function of the Si chemical potential
$\mu_{\mathrm{Si}}$ according to eq.~(\ref{eq:form}). Note that among those
structures with the same number of Si and Ag atoms, we only include the one
with largest adsorption energy per Si atom (see Table~\ref{eads}), since those
omitted run parallel in the plot but shifted upwards.
The origin for $\mu_{\mathrm{Si}}$ has been placed at the first crossing with
the formation energy of the clean Ag(110) surface (dark horizontal line)
--see 'Methods' for further details.
The shaded region indicates the most stable phase for
$\Delta\mu_{\mathrm{Si}}>0$, which under both XC functionals corresponds to
the P-MR DNR structure. 
The ZZ-MR and S-MR models, on the other hand, may be ruled out throughout
the entire  $\Delta\mu_{Si}$ range. Note that the P-MR SNRs
start to become more stable for $\mu_{Si}< 0$
in qualitative agreement with the
experimental observation, as SNRs are typically formed at small Si coverages,
while the DNR phase tends to cover the entire surface as the coverage is increased.}
\label{phasediag}
\end{figure*}

We follow the standard procedure of treating the Si chemical potential as a
parameter in eq.~(\ref{eq:form}) and plot the formation energies for each
structure as a function of $\mu_{Si}$ in Figures~\ref{phasediag}(a) and (b)
for the LDA and GGA derived energies, respectively. However,
since a reference structure for the Si reservoir is not available (and hence
the absolute value of $\mu_{Si}$ is unknown) we plot the formation
energies as a function
of a chemical potential shift, $\Delta\mu_{Si}$, whose origin is placed
at the first crossing between the formation energy of the clean Ag(110)
and that of any of the NRs (in our case it corresponds
to the P-MR DNR structure). Within this somewhat arbitrary choice,
small or negative values of $\Delta\mu_{Si}$ would correspond to Si poor conditions,
while large positive values to Si rich conditions. 

\putbib[nature]
\end{bibunit}

\newpage

\begin{table}[b]
\caption{{\bf Details of the P-MR/Ag(110) geometry:}
         Relative vertical distances
($z$) with respect to the average topmost Ag layer and bond distances to the first
silver nearest neighbor ($d_{Si-Ag}$) for each
of the symmetry inequivalent atoms in the SNR and DNR structures --see
Fig.~\ref{geom} for further details.
         \label{nn}}
\vskip 1.0 em
\begin{tabular}{ccccc}
\hline \hline
                     &  atom   & $z$ (\AA)& $d_{Si-Ag}$ (\AA)  \\
\hline
\multirow{2}{*}{SNR} & \Siad   &     1.42 &   2.56 ($\times2$), 2.80 \\
                     & \Sis    &     0.68 &   2.58, 2.74   \\
\hline
\multirow{4}{*}{DNR} & \Siado  &     1.44 &   2.55 ($\times2$), 2.87 \\
                     & \Siadt  &     1.38 &   2.56 ($\times2$), 2.78 \\
                     & \Siso   &     0.68 &   2.58, 2.73   \\
                     & \Sist   &     0.58 &   2.58, 2.76   \\
\hline \hline
\end{tabular}
\end{table}

\newpage

\newpage

\begin{table*}[b]
\caption{{\bf Si adsorption energies:}
         Adsorption energies per Si atom, in eV, obtained both under LDA and
GGA according to eq.(~\ref{eq:ads}) for the different MR models considered 
(except the S-MR). 
They range between $6.2-6.4$ for LDA and $5.5-5.7$~eV for GGA.
The large 0.7~eV difference between the two functionals is caused by their well
known over- and under-binding nature, respectively. Nevertheless,
both functionals point to the P-MR SNR and DNR as the most stable ones
(marked in bold face in the table) the two attaining very similar values.
         \label{eads}}
%\begin{tabular}{ccccc}
%\hline
% Structure   & $N_{Si}$ & $N_{Ag}$ & $E^{LDA}_{ads}$ & $E^{GGA}_{ads}$ \\
%\hline
%ZZ-MR SNR    &    4     &     88   &{\bf 6.36}       &   {\bf 5.59}    \\
%\hline
%ZZ-MR DNR1   &    8     &     86   &{\bf 6.35}       &   {\bf 5.58}    \\
%ZZ-MR DNR2   &    8     &     86   &     6.34        &        5.58     \\
%\hline
%P-MR SNR     &    6     &     88   &{\bf 6.44}       &   {\bf 5.70}    \\
%P2-MR SNR    &    6     &     88   &     6.24        &        5.56     \\
%\hline
%P-MR DNR     &   12     &     86   &{\bf 6.43}       &   {\bf 5.70}    \\
%P-MR DNR2    &   12     &     86   &     6.40        &        5.67     \\
%P2-MR DNR1   &   12     &     86   &     6.23        &        5.68     \\
%P2-MR DNR2   &   12     &     86   &     6.41        &        5.49     \\
%\hline \hline

\begin{tabular}{c|ccc|ccc|ccc}
\hline
Model&            \multicolumn{3}{c|}{SNR}       &    \multicolumn{3}{c|}{DNR1 ($-LL-RR-LL$)}  & \multicolumn{3}{c}{DNR2 ($-LR-RL-LR$)}  \\ \hline \hline
     &$N_{Si/Ag}$&$E^{LDA}_{ads}$&$E^{GGA}_{ads}$&$N_{Si/Ag}$&$E^{LDA}_{ads}$&$E^{GGA}_{ads}$&$N_{Si/Ag}$&$E^{LDA}_{ads}$&$E^{GGA}_{ads}$ \\ \hline
P-MR &     6/88  &   {\bf 6.44}  &   {\bf 5.70}  &    12/86  &  {\bf 6.43}   & {\bf 5.70}    &    12/86  &  6.40         &    6.57        \\ 
P2-MR&     6/88  &        6.24   &        5.56   &    12/86  &       6.23    &      5.68     &    12/86  &  6.41         &    5.49        \\ 
ZZ-MR&     4/88  &        6.36   &        5.59   &     8/86  &       6.35    &      5.58     &     8/86  &  6.34         &    5.58        \\ 
%S-MR &     4/90  &        0.00   &        0.00   &     8/90  &       0.00    &      0.00     &     8/90  &  0.00         &    0.00        \\ \hline
\end{tabular}
\end{table*}

\newpage

\end{document}